\documentclass[11pt,a4paper]{article}
\usepackage{color}
\usepackage[utf8]{inputenc}
\usepackage{lmodern}
\usepackage[onehalfspacing]{setspace}
\usepackage{amsmath}
\usepackage[scale=0.75]{geometry}
\usepackage{graphicx}
\usepackage{color}
\usepackage[small,bf, justification=justified]{caption}
\usepackage{wrapfig}
\usepackage[]{placeins}
\usepackage{hyperref}
\usepackage[backend=biber, style=numeric, sorting=none]{biblatex}
\usepackage{amssymb}
\usepackage{bbm}
\usepackage[gen]{eurosym}
\usepackage{tikz}
\usetikzlibrary{automata,positioning}
\usepackage{subfig}
\usepackage[T1]{fontenc}
\usepackage{algorithm,algpseudocode}
\usepackage[toc,page]{appendix}
\usepackage{caption}
\usepackage{cleveref}
\usepackage{siunitx}
\usepackage{ulem}
\usepackage{url}
\DeclareSIUnit\year{yr}

\bibliography{Literatur}

\title{The Characteristic Time Scale of Cultural Evolution}
\author{Tobias Wand$^{1,2*}$ \and Daniel Hoyer$^{3,4}$}
\date{%
    $^1$Westfälische Wilhelms-Universität Münster, Insitut für Theoretische Physik\\%
    $^2$Center for Nonlinear Science, Münster\\
    $^3$ George Brown College, Toronto \\
    $^4$ Evolution Institute, San Antonio\\
    $*$ Corresponding Author: \url{t\_wand01@uni-muenster.de}\\[2ex]%
    December 2022, Last Revision July 2023
}

\begin{document}

\maketitle

\begin{abstract}
Numerous researchers from various disciplines have explored commonalities and divergences in the evolution of complex social formations. Here, we explore whether there is a 'characteristic' time-course for the evolution of social complexity in a handful of different geographic areas. Data from the \textit{Seshat: Global History Databank} is shifted so that the overlapping time series can be fitted to a single logistic regression model for all 23 geographic areas under consideration. The resulting regression shows convincing out-of-sample predictions and its period of extensive growth in social complexity can be identified via bootstrapping as a time interval of roughly 2500 years. To analyse the endogenous growth of social complexity, each time series is restricted to a central time interval without major disruptions in cultural or institutional continuity and both approaches result in a similar logistic regression curve. Our results suggest that these different areas have indeed experienced a similar course in the their evolution of social complexity, but that this is a lengthy process involving both internal developments and external influences.
\end{abstract}


\section{Introduction}

\subsection{Motivation to Find Characteristic Time Scales}

Researchers from various disciplines have analysed commonalities and divergences in the evolution of complex social systems \cite{carballo_cooperation_2014,richerson_cultural_2013,Shin2020_PCA2,TurchinBook,Seshat_Tempo_Mode,Seshat_PC1,EvolutionaryDrivers}. The recent emergence of Cliodynamics as a discipline has started the analysis of the dynamics of human societies and states with data-driven scrutiny and modelling approaches from natural sciences \cite{turchin_arise_2008,manning_collaborative_2017}.  Previous work established that a common set of factors associated with complex social formations typically moved in tandem across a wide variety of regions and time-periods; factors such as social scale, the use of informational media,  administrative hierarchies, monetary instruments, and others \cite{Seshat_PC1}. These were interpreted as comprising the primary dimension of what could be called 'social complexity' across cultures, though other dimensions can be adduced as well \cite{Shin2020_PCA2}.\\

Various studies have already discussed or tried to identify the causal drivers of cultural evolution and evaluated the evidence for different theories of \textit{why} cultures become more complex \cite{EvolutionaryDrivers,Childe,White,Service,Kirch2010-vv}. Beyond the \textit{causal} similarities behind cultural evolution across cultures, researchers have also found evidence for \textit{temporal} similarities and seemingly parallel time scales in the dynamics of various social structures. For example, models for societal collapse have been derived from demographic and fiscal data that show characteristic oscillation periods of a few centuries and a fine structure with a faster periodicity of approximately two human generations \cite{TurchinBook}. Other theories suggest that cultural evolution leads to the emergence of similar political institutions and schools of thought at roughly identical time intervals across different geographic regions \cite{Spengler,Engels2018,Engels2021}. Another recent study has evaluated the connection between the first emergence of complex societies in different world regions and the age of widespread reliance on agriculture in those areas \cite{AgricultureToState}, supporting the theory that agriculture is a necessary condition for the evolution of complex societies. While the time lag between the primary reliance on agriculture and the emergence of states was found to decrease over time, an average time lag of roughly 3,400 years for pristine states suggested the existence of a characteristic time scale, though this was not the explicit focus of that study. Similarly, the study on causal drivers in \cite{EvolutionaryDrivers} also found that the time since the adoption of agriculture had a statistically significant effect as a linear predictor variable (called 'AgriLag') for sociopolitical complexity, providing additional evidence for temporal regularities in the growth of complexity across different cultures and civilisations. Finally, using the same data as our article (cf. section \ref{sec:Data}), it was possible to identify periods of cultural macroevolution with either slow or rapid change in social complexity \cite{Seshat_Tempo_Mode}. 



Nevertheless, as yet there is no consensus on whether there is a 'typical' time scale for socio-political development cross-culturally, let alone what that time-course might be. Such characteristic time scales of dynamic systems are, however, well documented in different areas of the natural sciences such as physics and chemistry \cite{Carr_2018,TimeScale_Chemistry}. Differentiating between fast and slow time scales in a dynamical system can lead to useful insights and can inform modelling assumptions for data analysis \cite{Separation_of_TimeScales,Willers2021}. In particular, Haken's theory of the 'enslaving principle' \cite{Haken1990-rf}, according to which the dynamics of fast-relaxing modes are dominated (enslaved) by the behaviour of slowly relaxing modes in a dynamical system, pioneered the research on how dynamics on different time scales influence each other in the same observed system. The existence of temporal regularities among societal dynamics would suggest that cultural evolution not only occurs in similar developmental stages across geographic regions and time periods, but also in similar time intervals. This would add an important dimension to our understanding of \textit{how} complex social formations evolve, and raise a number of critical questions about what drives these cross-cultural patterns. \\

Here, we adapt some of the methods employed in the natural sciences in an attempt to identify characteristic time scales in the evolution of complex societies. We utilise data collected by the \textit{Seshat: Global History Databank} \cite{Seshat2015,francois_macroscope_2016,turchin_introduction_2018}, a large repository of information about the dynamics of social complexity across world regions from the Neolithic to the early modern period \cite{Data_Continuous}. We find that, despite significant differences in the timing and intensity of major increases in social complexity reached by polities across the Seshat sample, there is a typical, quantitatively identifiable time course recognisable in the data. This result is robust to a variety of checks and covers polities from all major world regions and across thousands of years of history. Our findings offer a novel contribution to the study of cultural evolution, indicating the existence of a general, cross-cultural pattern in both the scale as well as the pace of social complexity development.

\subsection{Seshat Databank}
\label{sec:Data}

The \textit{Seshat: Global History Databank} includes systematically coded information on over 35 geographic areas and over 200 variables across up to 10,000 years in time steps of 100 years (\cite{Seshat2015,Data_Continuous}; see also publicly available data at \url{http://www.seshatdatabank.info/databrowser/}). During the time interval captured by the Seshat databank, these NGAs are occupied by over 370 different identifiable polities, defined as an "independent political unit". This sample is constructed by identifying all known polities that occupied part or all of each NGA over time (see \cite{Seshat2015,francois_macroscope_2016,turchin_introduction_2018} for details). The recorded variables are aggregated into nine complexity characteristics (CCs) and a principal component analysis shows that 77\% of the variation in the data can be explained by the first principal component (SPC1), which has almost equal contributions from all nine CCs \cite{Seshat_PC1}. In the case of missing data or expert disagreement in \cite{Seshat_PC1}, multiple imputation \cite{rubin2004multiple} was used to create several data sets with the differently imputed values which were aggregated into the principle component analysis. The NGAs in the Seshat data cover a wide geographical range and different levels of social complexity, though it is important to note that the Seshat sample is focused laregly on relatively complex, sedentary societies (but not exclusively). Data on the CCs is sampled at century intervals, giving a time series of each polity's estimated social complexity measure throughout its duration.\\

Seshat data has allowed researchers to quantitatively test hypotheses on cultural evolution such as identifying drivers of social complexity and predictors of change in military technology, for example gauging the effect of moralising religions on cultural evolution or predicting historical grain yields \cite{EvolutionaryDrivers,WarMachines,MoralizingReligions_2019,Agriculture2021}. Further analysis of the Seshat data includes a discussion of ideas from biological evolutionary theory with respect to the \textit{tempo} of cultural macroevolution, defined as "rates of change, including their acceleration and deceleration", concluding that "cultural macroevolution is characterized by periods of apparent stasis interspersed by rapid change" \cite{Seshat_Tempo_Mode}. These results strongly relate to the question of the present article, whether there is some generality in the time scale of cultural evolution in the Seshat data.

\subsection{Data on Culture/Polity Boundaries and Duration}
Each NGA's time series can contain data about very different polities that succeeded each other. Sometimes, a gradual and continuous change between the polities justifies treating predecessor and successor polities as closely related; for instance, in the Latium NGA (modern-day central Italy), Seshat records three separate polities for the Roman Republic, indicating the Early, Middle, and Late phases. These phases are culturally and (to a signfiicant degree) institutionally continuous, so can be treated as a single polity-sequence. In other cases, there may have been an invasion or mass migration as a clear break-point between the two polity's continuity; for instance, between the Ptolemaic Kingdom and Roman Principate polities in the Upper Egypt NGA. Data from \cite{Data_Continuous} and other information recorded in the Seshat sample, notably information on the relationship between polities, is here used to establish a list of continuous polities. 
The continuity is evaluated either as cultural continuity or as political-institutional continuity and our cutout data for both approaches is published on \cite{Data_WandHoyer}.

\subsection{Organisation of this Article}

Section \ref{sec:EDA} explains how we transformed the time series data on each NGA in the Seshat sample to establish a common reference point to investigate the time course of changes in social complexity across NGAs. In short, we shift each NGA's time series with respect to a single anchor time such that the transformed time variable \textit{RelTime} shows major overlap between the RelTime-vs-SPC1-curves of all NGAs. Exploratory data analysis for the whole dataset reveals that there is a logistic relationship between \textit{RelTime} and the SPC1 response variable. Section \ref{sc:TimeScales} identifies the time scale of growth from the lower to the upper plateau of the logistic curve via bootstrapping. The logistic curve is compared to a regression using only either the culturally or institutionally continuous time series and moreover, the duration of those continuous time series is compared to the estimated characteristic time scale. Finally, the results of the analyses are summarised and discussed in section \ref{sec:SummDisc}. The mathematical methods and technical details are discussed in the appendices \ref{sec:Methods} and \ref{app:Latium} and appendix \ref{app:Data} gives more details on the used data. 

\newpage
\section{Approach: Data Transformation and Exploratory Analyses}
\label{sec:EDA}

First, all raw SPC1 time series are rescaled via a min-max scaling, i.e.

\begin{equation}
    SPC1 = \frac{SPC1_{raw} - \min(SPC1_{raw}) }{\max(SPC1_{raw}) - \min(SPC1_{raw})}.
\end{equation}

This has the advantage of making the interpretation of high and low SPC1 values much easier as high/low correspond to close to 1 or close to 0, respectively. It also makes the parametrisation of a logistic curve easier by restricting the observed data to a range between 0 and 1.


\subsection{Anchor Time}
\begin{figure}[h]
    \centering
    \includegraphics[width = \textwidth]{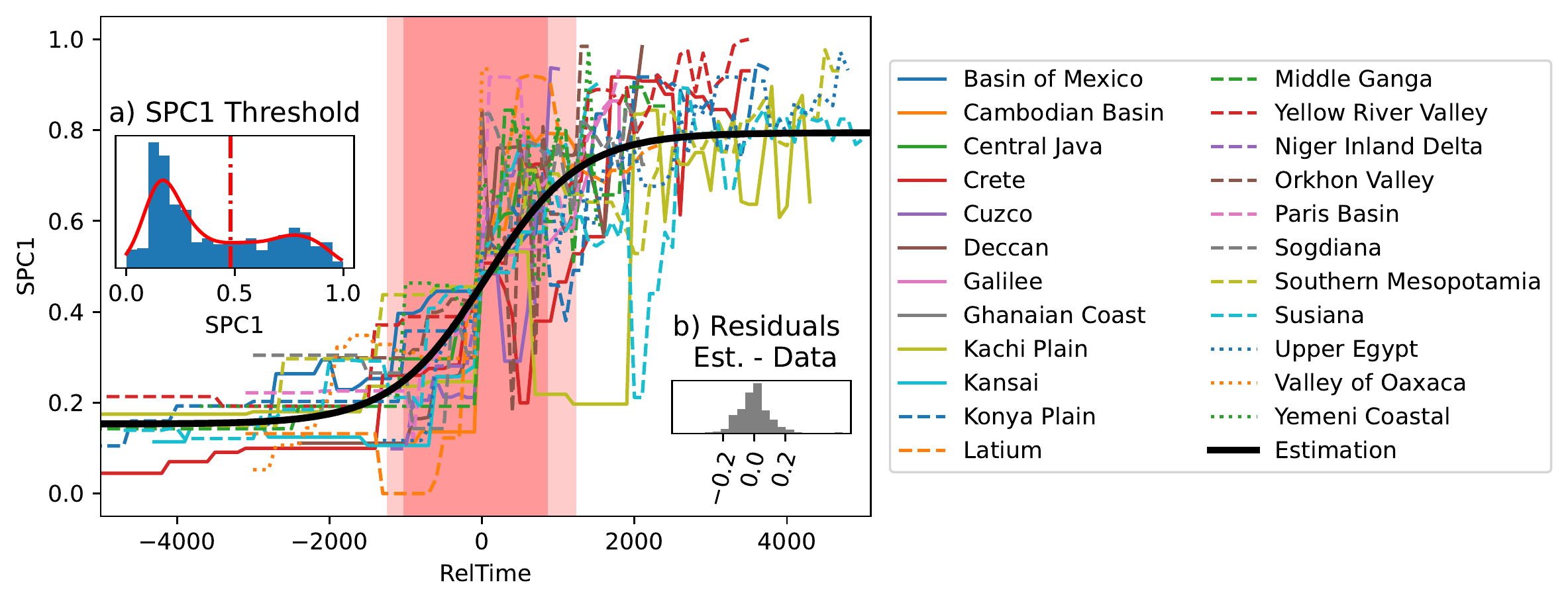}
     \caption{Main figure: Time series of RelTime vs. SPC1 for all 23 NGAs that cross $\textmd{SPC}1_0$ and the logistic regression. Marked in red is the area of growth between the two plateaus of the curve as identified in section \ref{sc:TimeScales}. The various time series are shown in appendix \ref{app:Data} in multiple plots to make the identification easier. Insets: a) distribution of SPC1 for all 35 NGAs, the associated KDE (red) and the threshold $\textmd{SPC}1_0$ (vertical); b) residuals of the logistic regression.}
    \label{fig:ExploratoryDataAnalysis}
\end{figure}

Considering that most NGAs have an SPC1 time series that starts at a low value barely above $0$ and ends at a high value close to $1$, a logistic regression model seems like a reasonable suggestion for the data. Although all NGAs experience a growth in SPC1 over time, they start at very different calendar years. Therefore, it is necessary to shift the time series via an anchor time so that in the new "relative" time, the growth phase in each NGA's time series coincide. Then, one logistic regression can be used for all shifted time series (cf. figure \ref{fig:ExploratoryDataAnalysis}; also \ref{method:LogisticFit}). Hence, each NGA $i$ needs an anchor time $T^{(i)}_a$ so that if all time series are shifted by $-T^{(i)}_a$, they roughly overlap. The shifted time series of the NGAs and the logistic fit are shown in the main part of figure \ref{fig:ExploratoryDataAnalysis}.

The anchor time can be chosen as the year during which the NGA $i\textmd{'s}$ SPC1 value crosses a threshold value. It has already been shown that there is a clear threshold $\textmd{SPC}1_0$ between high and low values of SPC1 in the data, which was used to define the \textit{RelTime} variable in \cite{MoralizingReligions_2019}. A similar methodology was also used in \cite{BigGods}, but there, the authors used the emergence of a moralising religious belief as the "year zero" to shift each NGA's time series. Copying the procedure from \cite{MoralizingReligions_2019} to get the \textit{RelTime} variable, $\textmd{SPC}1_0$ is chosen as the minimum between the two maxima in the kernel density estimation (KDE; explained in \ref{method:KDE}) of the SPC1 values (figure \ref{fig:ExploratoryDataAnalysis}, inset a). The anchor time $T^{(i)}_a$ is then selected as the first recorded data point when the NGA $i$ exceeds $\textmd{SPC}1_0$. An illustration of the anchor time shift is provided in the appendix \ref{app:Latium}. Thus, the 12 NGAs that never exceed $\textmd{SPC}1_0$ are discarded from this analysis. On the one hand, this is not too problematic because their limited growth in SPC1 means that they would have only contributed little information to the estimation of SPC1's characteristic growth time, but on the other hand, this discards all NGAs from the world region Oceania-Australia in the Seshat sample, meaning that it might introduce a bias. We discuss this and other possible limitations of the approach further in section \ref{sec:SummDisc} below.

\subsection{Logistic Regression}
The RelTime-vs-SPC1 data is fitted to a logistic regression curve (cf. \ref{method:LogisticFit}) via the optimisation algorithm \textit{scipy.optimize.curve\_fit} from \cite{2020SciPy-NMeth}. The quality of the regression curve is evaluated with the methods from \ref{method:RMSE}. With the exception of a few outlier observations occurring in several of the NGAs, all time series qualitatively agree with the regression curve fairly well. Also, the majority of the residuals shown in figure \ref{fig:ExploratoryDataAnalysis} (inset b) are distributed roughly symmetrically in a neighbourhood of zero. The distribution of the residuals and the rather low value of the root-mean-square-error $RMSE \approx 0.11$ both indicate that the logistic regression is a suitable model for the shifted SPC1 data.\\
To further increase our trust in the quality of the regression, it is evaluated via the coefficient of determination $\rho^2$ in an out-of-sample prediction. The data is split randomly into equally sized training and testing data sets and a logistic regression curve $f_i$ is estimated by only using the training data. Then, $f_i$ is used to predict the values for the test data and the prediction is evaluated via the $\rho^2$ metric in \ref{method:rho^2}. The random training-test-split is repeated $i=1,\dots,100$ times, each time using the estimated parameters from the full time series as initial values, and the resulting $\rho^2$ values have an average of $\rho^2 = 0.81 \pm 0.01$ far above 0 and therefore further strengthens our trust in the logistic model.

\section{Analysis of Time Scales}
\label{sc:TimeScales}
\subsection{Finding a Characteristic Time Scale}

Having established that the data can be accurately captured by a logistic curve, we can investigate our  research question; namely, how many years did it typically take in these different regions to transition from a polity with low SPC1 to one with high SPC1? Or to reformulate the question: when does the curve leave the low plateau and when does it reach the high plateau?  We attempt to answer these questions by estimating the heights of the plateaus and their respective uncertainties and by checking when the regression curve crosses these thresholds.\\
We performed 1000 steps of bootstrapping by sampling from the list of NGAs and by estimating the regression parameters $(a_i,b_i,c_i,d_i)_{i=1,\dots,1000}$ for each sample (cf. \ref{method:Bootstrapping}). According to the asymptotic behaviour in \ref{method:LogisticFit}, the plateaus are given by $b_i$ and $a_i+b_i$. In order to make conservative estimates instead of being influenced  by noise, an upper boundary for the lower plateau's value $Th_1$  and a lower boundary for the upper plateau's value $Th_2$  are used as the thresholds. $Th_1$ is chosen as $Th_1 = \mu(b) + 3\sigma(b)$ of the bootstrapped distribution of $b$, $Th_2$ as $Th_2 = \mu(a+b) - 3\sigma(a+b)$. For each bootstrapped logistic curve $f_i(t)$, it is then determined at which \textit{RelTime} values $t^{(i)}_1$ and $t^{(i)}_2$ it crosses the lower and upper thresholds $Th_1$ and $Th_2$. We can then understand the mean value 
\begin{equation}
\label{eq:characteristic_timescale}
    \mu\left(t^{(i)}_2-t^{(i)}_1\right) = \mu\left(t^{(i)}_2\right)-\mu\left(t^{(i)}_1\right)\approx  \SI{2500}{\year}
\end{equation}
as the characteristic time scale for the period of rapid cultural evolution between low and high plateaus of socio-political complexity, across geography and not in reference to any specific time period. Note that one can also choose less restrictive thresholds via $Th_1 = \mu(b) + \sigma(b)$ and $Th_2 = \mu(a+b) - \sigma(a+b)$. With these thresholds, the regression curve leaves the vicinity of the lower plateau rather quickly but needs much longer until it is close enough to the upper plateau to be considered as having reached the upper plateau. These $1\sigma$ thresholds would result in a longer time scale of roughly 

\begin{equation}
\label{eq:1-sigma-timescale}
    \mu\left(t^{(i)}_2-t^{(i)}_1  \middle| 1\sigma\right)\approx  \SI{4000}{\year}.
\end{equation}

We can check the general validity of these results by explicitly identifying for each NGA $i$ the first time $\tau_1^{(i)}$ that their SPC1 value exceeds $Th_1$ and the first time $\tau_2^{(i)}$ they exceed $Th_2$. With the exception of the Ghanaian Coast, all NGAs cross $Th_2$ and therefore, this procedure yields 22 time durations $d^{(i)} = \tau_2^{(i)}-\tau_1^{(i)}$. Only for the two NGAs Kachi Plain and Middle Yellow River Valley (two 'pristine' states, cf. the discussion in section \ref{sec:Discussion_TimeScale_Continuous}) does the duration $d^{(i)}$ exceed the 4000 years estimated as an upper boundary in \eqref{eq:1-sigma-timescale}. Both the mean (approximately 2200 years) and median (2100 years) are in line with the main estimation in \eqref{eq:characteristic_timescale}.

\subsection{Continuous Polities}
\begin{figure}[h]
    \centering
    \includegraphics[width = \textwidth]{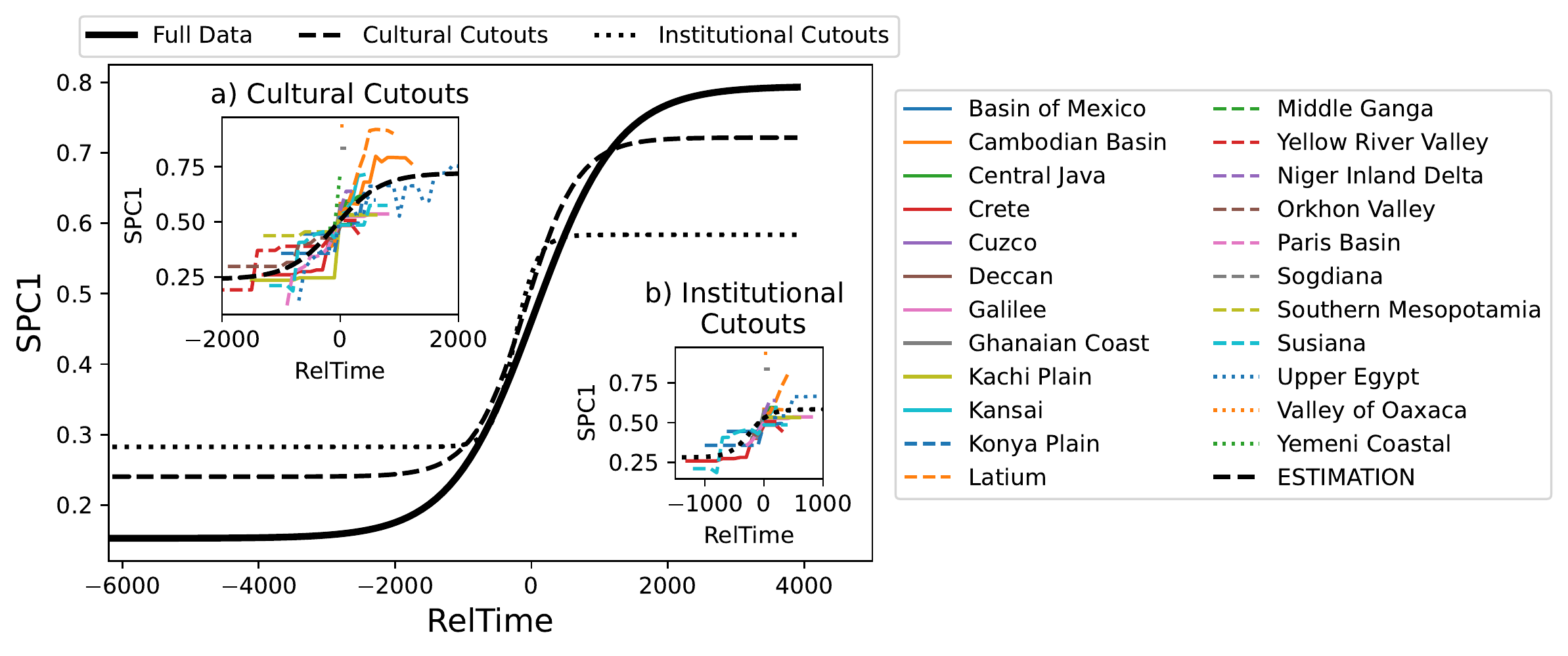}
    \caption{Main figure: estimated logistic curves for the full data and the two cutout methods. Insets: a) each NGA's central time series and resulting logistic curve for the culturally continuous time series; b) the same as subfigure a) for the institutionally continuous time series.}
    \label{fig:CompareCutouts}
\end{figure}

There are two reasons why it makes sense to restrict the logistic regression only to a central part of each NGA's time series, during which the polities in that NGA are not disrupted by external influence or major dislocations in socio-political structures. First, the logistic regression starts at a plateau of low values of SPC1 close to $0$ and ends at a plateau of high values close to $1$. Therefore, even a bad interpolation for the central part can achieve a good $RMSE$, if the plateaus of the high and low tails are sufficiently accurate. However, this would not be a reliable estimation to make an inference on the growth phase in the centre of the curve. Second, if the NGA's polity is e.g. annexed by another, more developed polity, then it inherits the invading polity's high SPC1 value and may make a sudden jump in the SPC1 curve. However, the logistic regression here is intended to model steady, uninterrupted growth like in \cite{Verhulst} and not major transitions driven by developments experienced elsewhere, as through annexations by an external invader. Therefore, it makes sense to divide each NGA's time series into intervals which are separated by sharp, discontinuous changes within each NGA and to restrict the analysis of the NGA to its central interval, i.e. to the time series from the polities that cross the $\textmd{SPC}1_0$ threshold.\\
As mentioned earlier, there are two ways of identifying such discontinuous changes: either via cultural changes of via major institutional changes of the polity's governance. Both approaches are analysed separately. The central time series for both methods and their resulting logistic regressions are shown in figure \ref{fig:CompareCutouts}.

\subsubsection{Cultural Continuity}

One set of sequences was determined by the absence of a major cultural dislocation; namely, the introduction of a new ideological and linguistic system, major population displacement, or major technological advance (the adoption of iron metallurgy, for instance). This is a very broad and lenient definition of continuity, as it allows for very different social formations to be part of a single sequence and can include significant developments. In Egypt, for instance, we treat nearly the entire Pharaonic period (from the Naqada period to the Achaemenid conquest) of over 3000 years, including the so-called Intermediate periods when central rule was fragmented (though many cultural and social features were retained), as a culturally continuous time period. \\
For the 23 NGAs under consideration, the mean value of data points for the culturally continuous central interval is approximately $11.7$, i.e. there is on average a bit more than one millennium of data. While this is much shorter than the characteristic time scale of roughly $2500$ years, the longest continuous time series of the NGAs show a similar length to that of the characteristic time scale (cf. the left half of table \ref{tab:Length_Continuity}). Hence, the logistic regression for these cutouts is rather close to the regression of the full data (cf. main part of figure \ref{fig:CompareCutouts}) and in particular, the regression curves' steepness (i.e. their time of growth) is quite similar.

\subsubsection{Institutional Continuity}

For institutionally continuous time periods, we follow a similar procedure as above, though with different criteria for continuity leading to shorter sequences. Namely, we break each sequence at any significant political/institutional change, even if there was much continuity in socio-cultural forms. In Egypt, for instance, the institutional sequence starts at the 1st Dynasty period and ends a the end of the Old Kingdom period and the First Intermediate Period, which we call the 'Period of the Regions'. The mean value of data points for institutionally continuous central time series is only $5.9$ and represents approximately 500 years of data. Even the longest continuous sequences now do not last as long as the characteristic growth time of $2500$ years (cf. right half of table \ref{tab:Length_Continuity}). Moreover, the logistic regression has only very little data for the parameter estimation (cf. inset \textit{b} of figure \ref{fig:CompareCutouts}) and hence, the logistic regression has a much lower SPC1 level for the upper plateau than the regression to the full data (main part of figure \ref{fig:CompareCutouts}), because the cutout time series are too short to reach the high-SPC1 levels.

\begin{table}[h!]
    \centering
    \begin{tabular}{c|c||c|c }
    NGA & Cultural Continuity Length & NGA & Institutional Continuity Length\\ \hline
       Yellow River  & 38  & Susiana  & 17 \\
       Upper Egypt & 33  & Crete & 17 \\
      Kachi Plain & 22 & Konya Plain & 15 \\
      Susiana & 21  & Upper Egypt & 10 
    \end{tabular}
    \caption{For both methods of identifying continuous time sequences, the four longest continuous central time series are shown and the amount of data points they contain (given as their length). The data points are sampled at intervals of one century. The culturally continuous time series are much longer than the longest institutionally continuous sequences.}
    \label{tab:Length_Continuity}
\end{table}

\FloatBarrier

\section{Summary and Discussion}
\label{sec:SummDisc}

\subsection{Summary}

Exploratory data analysis shows in figure \ref{fig:ExploratoryDataAnalysis} that the logistic regression is a suitable model for the RelTime-vs-SPC1 time series. Bootstrapping allows us to narrow down the time interval of rapid SPC1 growth to approximately 2500 years, as highlighted in figure \ref{fig:ExploratoryDataAnalysis}. Together, these results illustrate that there is a uniform behaviour in growth of social complexity represented by the time evolution of SPC1. \\
If the data is restricted to the central part of each NGA's time series without any discontinuous cultural or institutional transitions, the logistic regression is still a reasonable model and shows a similar shape to the full data as depicted in figure \ref{fig:CompareCutouts}. In particular, the regression based on the culturally continuous time series show a very similar steepness (i.e. growth period) to the full regression curve.

\subsection{Discussion of the Time Scale and Continuous Sequences}
\label{sec:Discussion_TimeScale_Continuous}
Figure \ref{fig:CompareCutouts} shows that the culturally continuous and institutionally continuous time series result in a similar logistic regression to the full data. Notably, table \ref{tab:Length_Continuity} shows that the culturally continuous time series have a much longer duration than the institutionally continuous ones. In particular, in the Yellow River Valley, Upper Egypt, Kachi Plain and Susiana, the culturally continuous time series is approximately as long (or even longer) as the characteristic time scale of SPC1 growth. This is expected for regions that saw the emergence of large, complex states relatively early in history and without any precedent from neighbouring societies – the so-called 'pristine' or 'primary' states \cite{Emergence_pristine_states,PrimaryStateFormation} – which these regions all experienced. However, this is not the case for most other NGAs, indicating that in those NGAs, the growth from the lower to the higher SPC1 plateau did not take place over the course of just one culturally continuous era, but rather included developments across cultural spheres and, in most cases, including developments being 'brought in from the outside' in the form of direct conquest or more indirect influence. The institutionally continuous time series are all significantly shorter than the characteristic growth time, as expected from the criteria used to generate those sequences. This is notable, as it suggests that in order to transition from low to high social complexity, major shifts in the NGA's governing institutions are necessary to facilitate the increase in social complexity. In other words, our findings suggest that major transitions in social complexity are not feasible for a single polity to accomplish, but require multiple social formations or 'phases' of rule building successively (but not monotonically, as the above figures illustrate) on prior developments. Nevertheless, the general similarity of the three regression curves in figure \ref{fig:CompareCutouts} shows that our analysis is stable with respect to the exact selection of time periods and different cutout criteria used. 

It is interesting to compare those NGAs that crossed the threshold $\textmd{SPC}1_0$ to those that failed to do so and stayed at lower complexity values. The latter group had a mean of only 6.4 recorded data points, i.e. there were only complex social formations coded as part of the Seshat sample for a period of roughly six centuries. On the other hand, the NGAs that did reach a high complexity and exceeded the threshold $\textmd{SPC}1_0$ had a mean of 57.3 recorded data points, corresponding to almost 6 millennia of observed data. Partly this is explained by different availability of historical and archaeological evidence in different regions, but it suggests also that cultural developments in the low complexity NGAs could have followed the same trajectory of logistic growth, if they had been given enough time.
Unfortunately, the necessity to identify an anchor time for this analysis means that all NGAs from the Seshat world region Oceania-Australia had to be discarded for this research. The bias introduced by this has to be kept in mind while interpreting our results.

\subsection{Interpretation and Comparison to Previous Work}

With the shifted time index RelTime, the logistic regression model of the SPC1 time series achieves a high accuracy in capturing the evolution of socio-political complexity measured by SPC1. Previous work has already demonstrated a significant amount of cross-cultural generality in the factors contributing to the evolution of socio-political complexity (\cite{Seshat_PC1}, supplemented by findings in \cite{Shin2020_PCA2,kohler2022social}). Notably, a previous study has already identified a characteristic growth pattern of SPC1 and the second principal component SPC2 and found that a rapid period of scale is first followed by a growth of information processing and economic complexity and then by further growth in scale \cite{Shin2020_PCA2}. \\
Here, we expand on this prior work by identifying that the time scales involved in these developments also exhibit a general, characteristic shape. Nevertheless, the evolution of social complexity is a lengthy and non-monotonous process; this emerges clearly from our analyses distinguishing the full regional time-series involved in the transition from low to high thresholds of SPC1 from sequences of cultural or institutional continuity. We see no examples of this evolution accomplished during a single institutionally-continuous sequence. Further, in all NGAs there are noisy periods during which SPC1 grows but also crises during which socio-political complexity sharply declines, only to recover later and continue increasing. These findings highlight both that different parts of the world experienced similar processes of social complexity growth, involving multiple phases of cultural and socio-political structures building off of (and occasionally recovering from) prior developments in each region. \\
While the sample of past societies explored in this article is certainly not exhaustive, they comprise a fairly representative sample of regions from different parts of the world and include societies from different periods, cultures and different developmental experiences. Our results thus lends novel empirical support to the idea (from e.g. \cite{Spengler,Seshat_Tempo_Mode}) that socio-cultural evolution does indeed occur in similar time scales across different cultures and geographies. Future research can expand these insights by including additional societies and exploring alternate thresholds of complexity to identify anchor times to include more NGAs from the original sample, because the current thresholding procedure in particular excluded some NGAs from modern-day Oceania-Australia from our analysis.\\
In terms of the underlying approach, our study tries to single out the autocatalytic effect of social complexity growth. To this end, it not only focused on one NGA at a time, but also compares our regression results to the culturally and institutionally continuous periods for the respective NGA. Thus, we uncover an empirical pattern in the temporal evolution of SPC1 that has not yet been fully discussed by previous work, e.g in causal analyses of the drivers of social complexity like in \cite{EvolutionaryDrivers}. Our methodology differs from e.g. the regression model in \cite{EvolutionaryDrivers} by deliberately choosing a very simple model to single out the temporal evolution whilst disregarding possible drivers of the observed dynamics. We believe this approach can be utilised to answer other questions about long-run cultural evolution, for instance the processes by which key technologies (e.g. metallurgy, military technology, communications media etc.) are invented in certain locations and then adopted in others.  \\
While the autocatalytic growth model provides an elegant interpretation of our findings (the current level of complexity facilitates further growth until the presence of an upper boundary of complexity is approached), it has to be regarded with caution: We sought as far as possible to disentangle culturally and institutionally endogenous developments from those driven by interactions with other polities, though even the internal developments are not free from external influence. Previous work, for instance, shows the strong effect of military conflicts with other states on the growth of sociopolitical complexity \cite{EvolutionaryDrivers,WarMachines}. Hence, the autocatalytic model might be a useful low-dimensional description of the data, but not an exhaustive explanation. In short, our findings exposes a cross-cultural temporal pattern whose causes need to be fleshed out in future work. \\
Finally, the findings of the present article can be used as a benchmark for future additions to the Seshat data: if a new NGA is added to the databank and shows a clear divergence from the logistic curve, it may be prudent to either check, if there are any mistakes in the data generation and interpolation, or if the divergences can be explained by historical developments. Such a benchmark may thus be useful for further expansion of the Seshat databank.


\subsubsection*{Author Contributions}
TW performed all analyses and drafted the manuscript; DH assisted in conceptual development and drafting the manuscript.

\subsubsection*{Acknowledgements}
Initial ideas behind this paper were developed at a workshop held by the Complexity \& Collapse Research Group of the Complexity Science Hub, Vienna. The authors thank all the participants at that event, particularly Mateusz Iskrzyński for valuable contributions at early stages of this project. Financial support for this work was provided by the “Complexity Science” research initiative supported by the Austrian Research Promotion Agency FFG under grant \#873927 and by the German Academic Scholarship Foundation (Studienstiftung des deutschen Volkes).

\printbibliography

\FloatBarrier

\clearpage

\begin{appendices}
\section{Methods and Technical Details}
\label{sec:Methods}

\subsection{Logistic Regression Curve}
\label{method:LogisticFit}
Logistic regression is used to model time series data which is mostly distributed at two plateaus with a transitory area between them \cite{LogisticRegression}. It is based on the characteristic sigmoid curve of the logistic growth model described in \cite{Verhulst}, which models an exponential growth process constrained by a carrying capacity. The logistic curve has the functional form $f$ with an asymptotic behaviour
\begin{equation}
\label{eq:LogisticCurve}
    f(x) =  \frac{a}{1 + \exp(-c (x - d))} + b,\quad  f(-\infty) = b \quad \textmd{ and }\quad f(\infty) = a+b.
\end{equation}
Often, data is scaled such that $b=0$ and $a = 1$, i.e. an asymptotic behaviour between two binary plateaus at height $0$ and $1$.

\subsubsection{Reversing the Direction}
\label{method:ReversionOfLogistic}
Estimating the coefficients $(a,b,c,d)$ can lead to numerical instabilities because it is possible to transform a logistic curve with $c>0$ to an equivalent equation $\hat{f}$ with $c<0$. Consider e.g. $a = 1, b=0,c=1$ and $d=0$, then
\begin{equation}
    f(x) = \frac{1}{1+\exp(-x)}  = \frac{\exp(x)}{\exp(x)+1} =  \frac{\exp(x) +1-1}{\exp(x)+1} = 1 + \frac{-1}{1+\exp(x)}.  
\end{equation} 
The last reformulation of $f$ can now be parametrised via $\hat{a}=-1, \hat{b} = 1, \hat{c}=-1$ and $\hat{d}=0$. This ambiguity can lead to the regression algorithm yielding positive and negative results for $c$ during multiple runs. This can be prevented by setting an initial parameter guess with $c>0$, which locks the algorithm into positive values for $c$.

\subsection{Kernel Density Estimation (KDE)}
\label{method:KDE}
A KDE tries to reconstruct a probability density function based on a sample $x_1,\dots,x_n$ of measurement data by smoothing the histogram of the data \cite{Parzen1962,Rosenblatt1956}. The estimated density $\hat{\rho}(x)$ is modelled as a weighted sum of probability densities (kernels) centred around the measured $x_i$. In this article, the Gaussian density is used as the kernel via \textit{scipy.stats.gaussian\_kde} \cite{2020SciPy-NMeth}.

\subsection{Residuals and Root Mean Squared Error}
\label{method:RMSE}

For an algorithm $f$ which estimates values $\hat{y}$ from data $X$ with true values $y$, there are several methods to evaluate the accuracy of $f$. One of them is the root mean squared error $RMSE$. It is defined as 
\begin{equation}
    RMSE = \sqrt{\frac{1}{n} \sum_{i=1}^n r_i^2 }
\end{equation}
via the residuals $r_i = \hat{y}_i -y_i$. An $RMSE$ much smaller than the range of measured values $y_i$ means that the model shows only little deviation from the data. A roughly symmetric distribution of the residuals around $0$ indicates that the model does not have a bias towards particular values.

\subsection{Coefficient of Prediction $\rho^2$}
\label{method:rho^2}
Another method to evaluate the quality of an estimated function $f$ is the coefficient of prediction $\rho^2$ used in \cite{Seshat_PC1}. It takes the value of $\rho^2=1$, if the prediction is always exactly true, and $\rho^2 = 0$, if the prediction is only as accurate as always using the mean $\Bar{y}$. It is defined by 
\begin{equation}
    \rho^2 = 1 - \frac{  \sum_{i=1}^n (\hat{y}_i - y_i)^2  }{ \sum_{i=1}^n(\Bar{y} - y_i)^2 }.
\end{equation}

\subsection{Bootstrapping}
\label{method:Bootstrapping}
Bootstrapping is used to estimate standard deviations and confidence intervals in a model-free approach. A sample $z_1, \dots,z_n$ is re-sampled \textit{with} replacement, i.e. a new sample $\Tilde{Z} = z_{i_1}, \dots, z_{i_n}$ is created that for some $j\neq k$ fulfils $i_j = i_k$. This procedure is repeated $N$ times so that there are $\Tilde{Z}_1, \dots, \Tilde{Z}_N$ bootstrapped samples. If $N$ is large enough, then e.g. the mean $\Tilde{\mu}(z)$ of the re-sampled data will converge to the true mean of the original sample, but the empirical distribution of the re-sampled means $\Tilde{\mu}_1(z),\dots,\Tilde{\mu}_N(z)$ enables the calculation of the confidence interval of the empirical mean \cite{Efron1993}. This approach can be adapted to make inference on the standard deviation and CIs of any statistical property of the original sample.

\clearpage

\section{Example of the Data Preprocessing}
\label{app:Latium}

\begin{figure}[h!]
\centering
\includegraphics{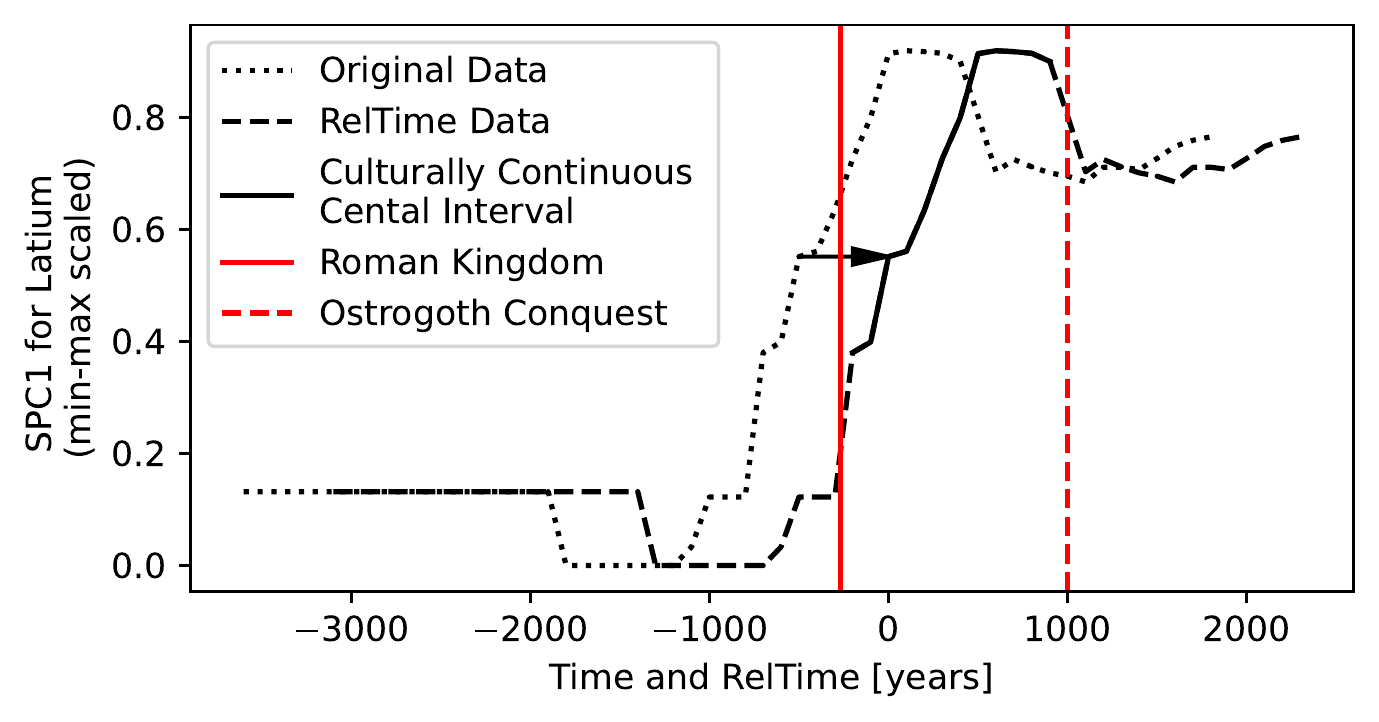}
    \caption{Illustration of how the $\textmd{SPC}1$ time series of the Latium NGA is shifted by its anchor time to the new relative time frame $RelTime$. Hence, at $RelTime=0$, the shifted time series is equal to the threshold value $\textmd{SPC}1_0$. The two recorded discontinuities for the Latium NGA (red) are used to divide its $\textmd{SPC}1$ time series into different intervals. The interval containing the threshold value $\textmd{SPC}1_0$ (the "Relevant Snippet" with the solid line) is used for further analysis whereas the rest (dotted) is discarded. Note that the times of the discontinuities do not line up with the 100-year time intervals of the $\textmd{SPC}1$ measurement which is why the transition between the black dashed/solid lines does not perfectly align with the red vertical lines.}
    \label{fig:Latium}
\end{figure}

We illustrate the preprocessing of the raw $\textmd{SPC}1$ time series using the NGA 'Latium' (modern day central Italy) as an example. Figure \ref{fig:Latium} shows the shift between the original time series to the \textit{RelTime} time frame relative to the anchor time $T^{(Latium)}_a = 500 \textmd{ BC}$. Additionally, this figure depicts how the $\textmd{SPC}1$ time series of Latium is dissected into culturally or institutionally continuous time series intervals of SPC values for the NGA. Note that the time index is given in $RelTime$ for the shifted data and real-world time for the original data.

\clearpage

\FloatBarrier

\section{Detailed Data}

\begin{figure}[h!]
    \centering
    \includegraphics[width = \textwidth]{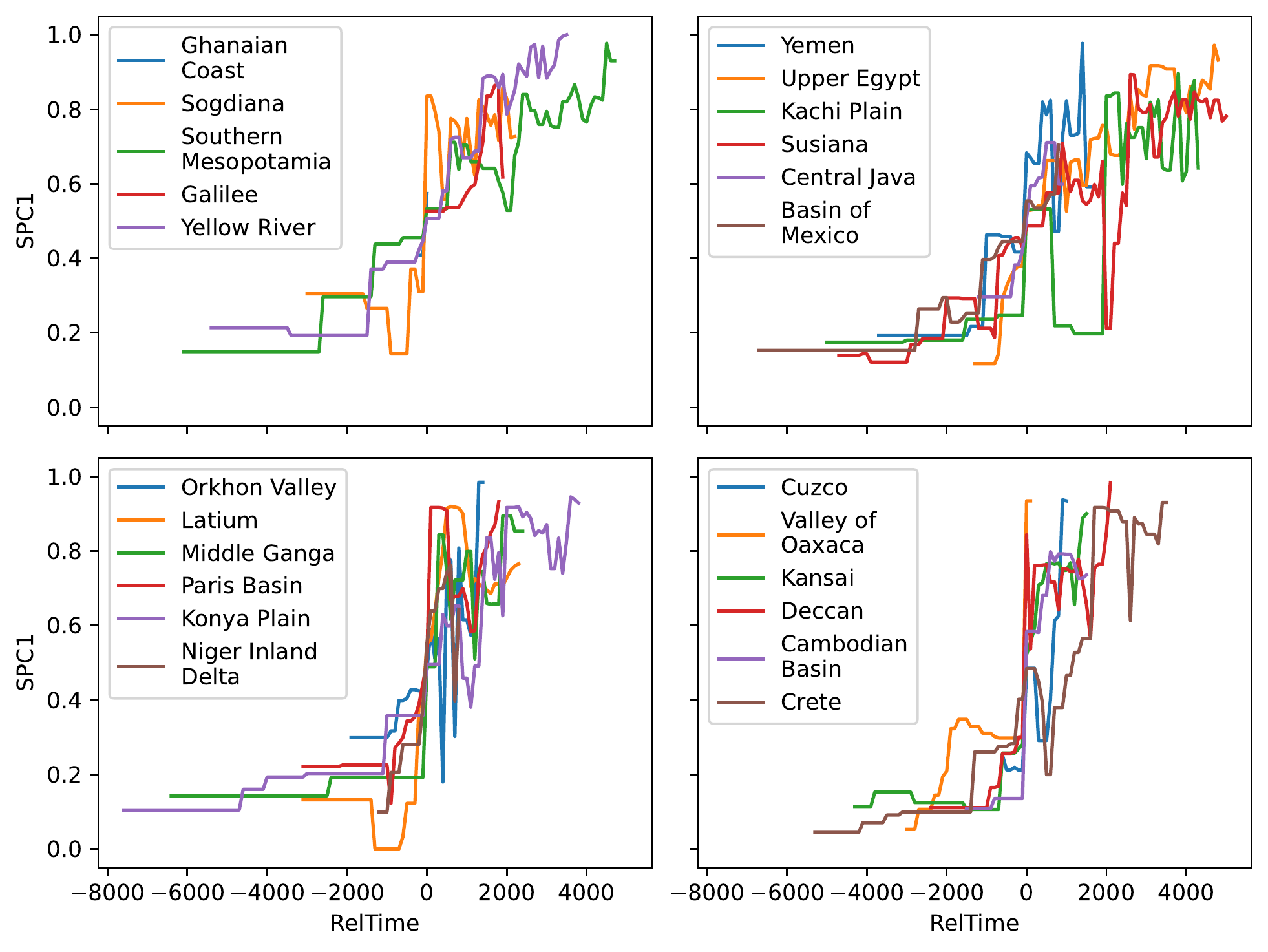}
    \caption{Spreading the various time series from figure \ref{fig:ExploratoryDataAnalysis} onto multiple plots to make it easier to distinguish the NGAs from each other.}
    \label{fig:Multiplot}
\end{figure}

The SPC1 data from figure \ref{fig:ExploratoryDataAnalysis} is shown in figure \ref{fig:Multiplot}, but spread out onto several subplots to help the reader identify different NGAs.

\label{app:Data}

\newpage
\subsection{Data Collection and Availability}

As described in the main text, the data used in this paper is derived from \cite{Data_Continuous}, supplemented with information provided by the authors. The original data is described in \cite{EquinoxData}. It was gathered, cleaned, reviewed, and managed by members of the Seshat Databank project following standard project methods, as described in the references cited in the main text. For more details on the social complexity data utilised here and development of the SPC1 values, see especially \cite{Seshat_PC1} and \cite{EvolutionaryDrivers}.\\
The full dataset used in the analyses presented here is available on \cite{Data_WandHoyer}. This shows:
\begin{itemize}
    \item \textbf{NGA}: Name of the NGA 
    \item \textbf{PolID}: Unique identifier for each polity in the sample
    \item \textbf{AbsTime}: The 'absolute' or calendar time-point for each SPC1 value. Note that we sample at 100-year intervals, for as many intervals as there are polities in the sample occupying each NGA 
    \item \textbf{RelTime}: The shifted time-series, such that RelTime = 0 in the century interval in each NGA during which SPC1 values crossed the calculated threshold, as described in the main text. Other rows in the NGA are express their relation before or after this threshold, in century intervals. Note that rows that fall outside of the central cultural or institutional sequence are not expressed  
    \item \textbf{SPC1}: Raw SPC1 values for each polity at each century interval, calculated as described in the main text (but not yet min-max scaled)
    \item \textbf{Culture.Sequence}: Label indicating the central time-series identified as part of the culturally-continuous sequence that surrounds each NGA's RelTime=0 threshold (labelled 'cultural.continuity'); all other century intervals in the NGA are labelled 'outside.central' to indicate they fall outside of this central interval sequence
    \item \textbf{Institutions.Sequence}: Same as above, but indicating the institutionally continuous sequence. 
\end{itemize}

\end{appendices}
\end{document}